\documentclass[aip,jcp]{revtex4-1}
\usepackage{amssymb,amsmath}

\newif\ifpdf
\ifx\pdfoutput\undefined
\pdffalse % we are not running PDFLaTeX
\else
\pdfoutput=1 % we are running PDFLaTeX
\pdftrue
\fi

\ifpdf
\usepackage[pdftex]{graphicx}
\else
\usepackage{graphicx}
\fi

\begin{document}

\ifpdf
\DeclareGraphicsExtensions{.pdf, .jpg, .tif}
\else
\DeclareGraphicsExtensions{.eps, .jpg}
\fi

\title{High-field Overhauser DNP in silicon below the metal-insulator transition}
\author{Anatoly E. Dementyev}
\affiliation{Francis Bitter Magnet Laboratory,
Massachusetts Institute of Technology, Cambridge, MA 02139}
%\email[]{}
%\homepage[]{}
%\thanks{}
\author{David G. Cory}
\affiliation{Institute for Quantum Computing and Department of Chemistry, University of Waterloo, Waterloo ON N2L 3G1, Canada}
\affiliation{Perimeter Institute for Theoretical Physics, Waterloo, ON N2L 2Y5, Canada}
\author{Chandrasekhar Ramanathan}
\email[Please address correspondence to: ]{chandrasekhar.ramanathan@dartmouth.edu}
\affiliation{Department of Physics and Astronomy, Dartmouth College, Hanover NH 03755}

%\affiliation{Department of Nuclear Science and Engineering,
%Massaschusetts Institute of Technology, Cambridge, MA 02139}

\date{\today}

\begin{abstract}

\begin{center}
{\large Abstract}
\end{center}
\noindent Single crystal silicon is an excellent system in which to explore dynamic nuclear polarization (DNP), as it exhibits a continuum of properties from metallic to insulating as a function of doping concentration and temperature.   At low doping concentrations DNP has been observed to occur via the solid effect, while at very high doping concentrations an Overhauser mechanism is responsible.  Here we report the hyperpolarization of $^{29}$Si in n-doped silicon crystals, with doping concentrations in the range of 1--3$\times 10^{17}$ cm$^{-3}$.  In this regime exchange interactions between donors become extremely important.  The sign of the enhancement in our experiments and its frequency dependence suggest that the $^{29}$Si spins are directly polarized by donor electrons via an Overhauser mechanism within exchange-coupled donor clusters.  The exchange interaction between donors only needs to be larger than the silicon hyperfine interaction (typically much smaller than the donor hyperfine coupling) to enable this Overhauser mechanism.  Nuclear polarization enhancement is observed for a range of donor clusters in which the exchange energy is comparable to the donor hyperfine interaction.  The DNP dynamics are characterized by a single exponential time constant that depends on the microwave power, indicating that the Overhauser mechanism is the rate-limiting step.  Since only about 2\% of the silicon nuclei are located within one Bohr radius of the donor electron, nuclear spin diffusion is important in transferring the polarization to all the spins.  However, the spin-diffusion time is much shorter than the Overhauser time due to the relatively weak silicon hyperfine coupling strength.  In a 2.35 T magnetic field at 1.1 K, we observed a DNP enhancement of $244 \pm 84$ resulting in a silicon polarization of $10.4 \pm 3.4$ \% following two hours of microwave irradiation.
\end{abstract}

%\pacs{}
%\keywords{}	% not needed by APS

\maketitle

\section{Introduction}

\noindent Microwave-induced dynamic nuclear polarization (DNP) occurs in a wide range of systems including metals, semiconductors and insulators, both in the bulk as well as in engineered nanostructures.  The physical mechanisms underlying DNP can be quite different in each case.  Single crystal semiconductors such as silicon are an excellent system in which to explore DNP, as they can exhibit a continuum of properties between those of metals and insulators depending on the doping concentration and temperature \cite{Abragam-1958,Jerome-1965,Henstra-1988,Dyakonov-1992, Fasol-2002a,Hayashi-2006,Hayashi-2009}.  

The electron spin resonance properties of shallow donors such as phosphorus and arsenic in silicon have been studied extensively \cite{Feher-1959}.  At low temperatures and low  doping concentrations the donor electrons are localized at the individual donor sites.  These sites are sparsely distributed, and essentially isolated from each other.  The sample is an insulator, and its ESR spectrum shows well resolved hyperfine splittings with the donor nucleus \cite{Feher-1959}.   There are about 1700 silicon nuclei located within the $\approx 2$ nm electron Bohr radius of the shallow donors { (without the donor-dependent central-cell correction)}, corresponding to about   
80--100 $^{29}$Si nuclei in a natural abundance sample.
The hyperfine couplings to these silicon nuclei are significantly weaker and show up as an inhomogeneous broadening of the ESR lines \cite{Feher-1959}.  DNP of silicon nuclei has been observed following microwave irradiation under these conditions, and is driven by the solid effect \cite{Abragam-1961}.  The presence of anisotropic hyperfine interactions admixes the nuclear spin eigenstates, enabling us to drive nominally forbidden transitions.  DNP is mediated by the silicon nuclei closest to the donor that have both the largest hyperfine interaction strengths and anisotropies.  The bulk nuclear spins are polarized by spin diffusion from these sites \cite{Ramanathan-2008}.  The DNP enhancements observed in this regime have typically been quite low, as the electron spin T$_1$ in silicon becomes extremely long at low temperatures.  Hayashi {\em et al.}\  have tried to improve DNP efficiency in this regime using optically excited carriers to shorten the electron spin T$_1$, with some success \cite{Hayashi-2009}.

As the donor concentration is increased the average distance between donor sites decreases and the strength of the exchange interaction between the donor electrons becomes much stronger.  At high doping concentrations, the ESR spectrum collapses to a single exchange-narrowed line before the system undergoes a metal-insulator transition \cite{Feher-1959}.  (The critical concentrations for phosphorus- and antimony-doped silicon are in the region of 3--4$\times 10^{18}$ cm$^{-3} \;$ \cite{Ochiai-1975}.)  The DNP in this case is mediated by the Overhauser effect, where the fluctuating contact hyperfine interaction leads to electron-nuclear cross-relaxation processes \cite{Overhauser-1953}.  Here the delocalized electrons have direct Fermi contact interactions with all the silicon nuclear spins in the sample, and DNP occurs at all the nuclear spin sites.  Silicon DNP has been observed in this high-doping regime as well, though here the electron spin T$_1$ is so short that it is very difficult to saturate the ESR transitions and achieve large DNP enhancements \cite{Jerome-1965}. 

Intermediate between these two regimes, the properties of the system change dramatically as a function of the doping concentration, as exchange processes are turned on.  At liquid helium temperatures the electron spin T$_1$ in phosphorus-doped silicon changes by 8 orders of magnitude  as the doping concentration is varied from $5 \times 10^{16}$ cm$^{-3}$ to $6 \times 10^{17}$ cm$^{-3} \;$ \cite{Cullis-1975}.  
{ Hayashi {\em et al}.\  recently observed an Overhauser enhancement  for a phosphorus-doped natural abundance silicon sample (N$_D = 10^{17}$ cm$^{-3}$) at 12 K at low magnetic fields \cite{Hayashi-2009}.  They also found that the spin-diffusion process was not a rate-limiting step in their DNP experiments on natural abundance silicon for doping concentrations in the range N$_D$  =  $10^{15}-10^{17}$ cm$^{-3}$.  This is inspite of the fact that at most about 2\% of the silicon nuclei are located within a Bohr radius of the donor electron in this doping range. 

In this paper we report on the high-field DNP enhancement of silicon nuclei in the intermediate-doping regime, and show that at high magnetic fields the Overhauser enhancement of the silicon nuclei occurs within exchange-coupled donor clusters. This Overhauser-type enhancement is observed even though the sample remains insulating at the temperatures used, and the exchange interactions are not strong enough to exchange-narrow the ESR line.   At high fields the nuclear Zeeman interaction frequently dominates the hyperfine interaction, while at low fields the hyperfine interactions are often larger than, or comparable to the nuclear Zeeman interactions.  The relative strengths of these interactions is important, as it changes the effective quantization axis of the nuclear spins, and the resulting selection rules for microwave irradiation.  This is one reason why higher DNP enhancements are often observed at low magnetic fields compared to higher fields.  High field dynamic nuclear polarization has been investigated both for the improved chemical sensititivity available at high magnetic fields \cite{Maly-2008}, as well as for the ability to create highly polarized nuclear spins states.  These highly polarized nuclear spins states are of interest in studying nuclear spin ordering \cite{Abragam-1982}, as well as for the preparation of pure quantum states \cite{Cory-2000}.  In order for the initial thermal polarization of the electron spins to be high, it is necessary to use field strengths of a few Tesla even at liquid helium temperatures.  This is the regime in which the experiments presented in this paper have been performed, and where we have achieved a silicon polarization of $10.4\pm3.4$ \%, which we believe is the highest that has been reported to date.}

\begin{figure}
    \includegraphics[width=5in]{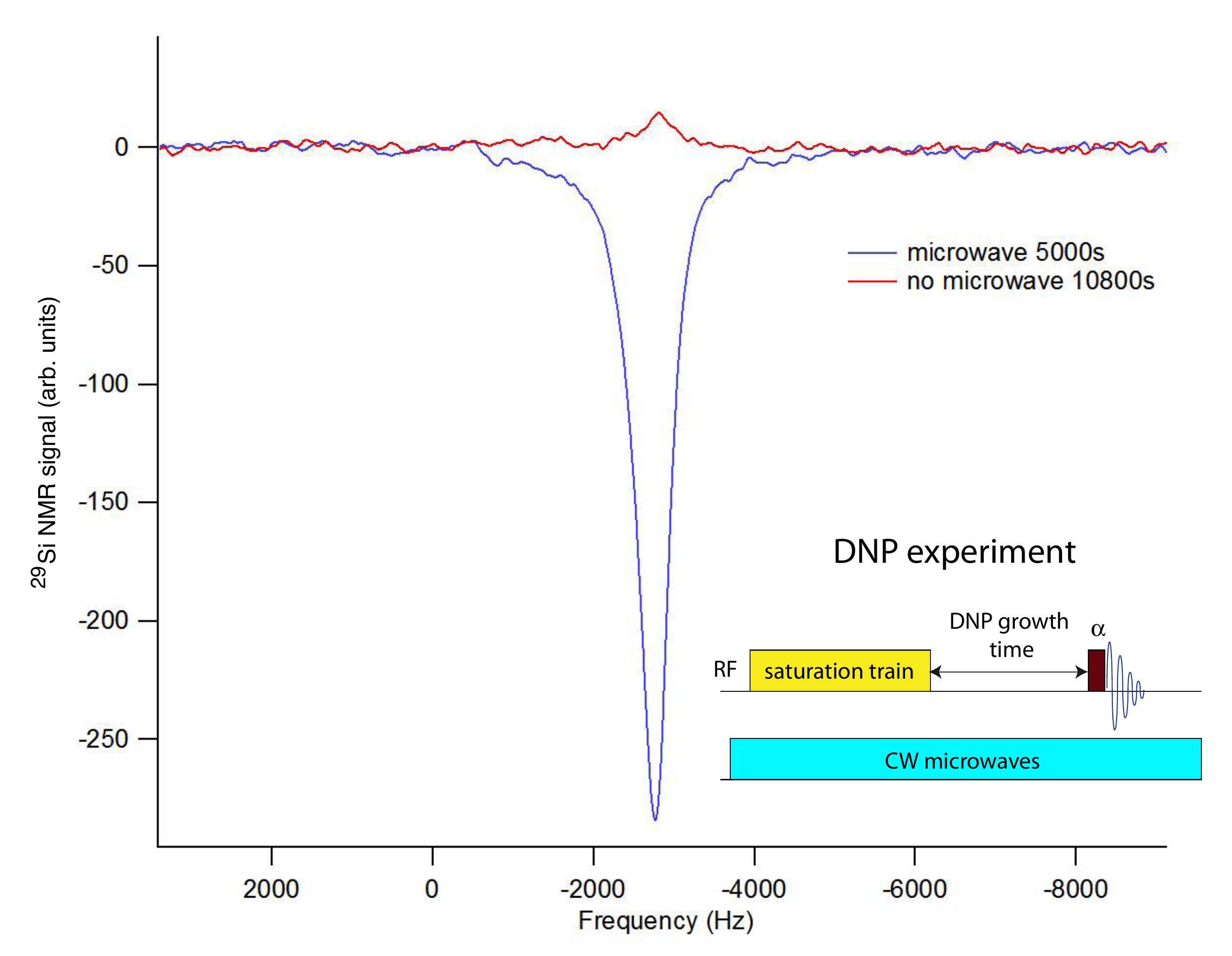} 
    \caption{DNP enhancement of 20 obtained following microwave irradiation of an antimony-doped silicon wafer ($2.5\times10^{17}$ cm$^{-3}$) at 1.4 K.  Note that the hyperpolarized signal is opposite to the thermally polarized signal.   The inset shows the basic DNP experiment. (color online)}
    \label{fig:enhancement}
\end{figure}

\begin{figure}
    \includegraphics[width=5in]{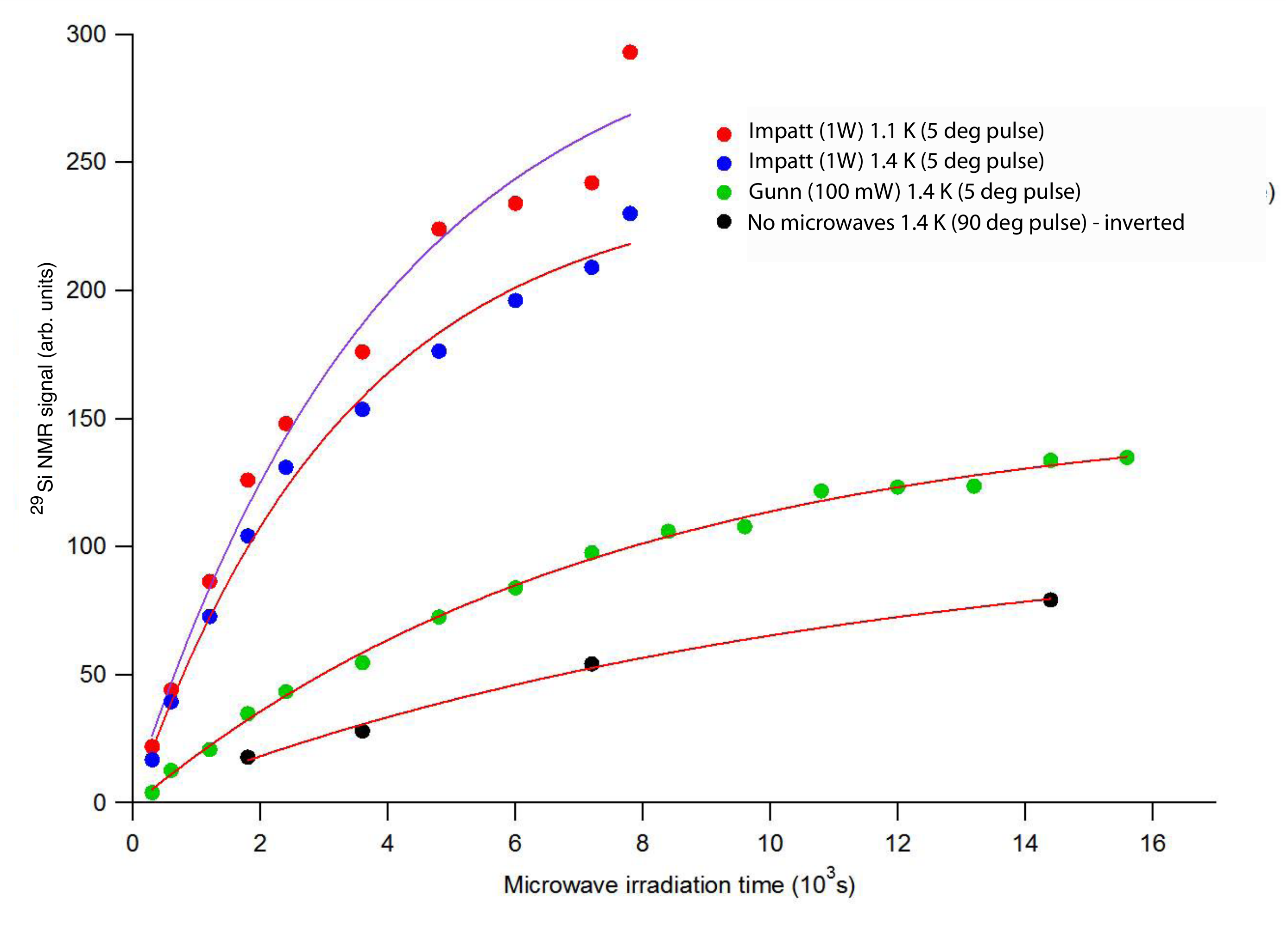}
       \caption{Build-up  of $^{29}$Si polarization in antimony-doped silicon ($2.5\times10^{17}$ cm$^{-3}$).  The enhancement obtained with the 100 mW source is clearly power-limited.  With the 1 W microwave source the hold time of the cryostat shortened from 4 hours to 2 hours.  A 5 $\deg$ pulse was used to monitor the polarization in the DNP experiments while a 90 $\deg$ pulse was used in the thermal equilibrium situation (color online).}
   \label{fig:power}
\end{figure}

\section{Results and Discussion}
\subsection{DNP Enhancement}
\noindent Figure \ref{fig:enhancement} shows the result of a DNP experiment using an antimony-doped sample. 
The experimental scheme is shown in the inset of the figure.  We apply a saturation train of $\pi/2$ pulses to initially destroy the nuclear spin polarization.  Following a variable delay during which the nuclear spin signal grows due to spin-lattice relaxation (if the microwaves are off) or DNP (if the microwaves are on), the polarization is monitored by a single nutation pulse (denoted by $\alpha$).   The experiments were performed at 2.35 T at a temperature of 1.4 K.  The thermal electron spin polarization is 81 \% while the thermal $^{29}$Si polarization is 0.034 \% under these conditions.  We used a 100 mW Gunn diode source (Millitech) tuned to 66 GHz, coupled to a standard microwave gain horn to drive the electron spin transitions in the DNP experiment \cite{Cho-2007}.   The figure shows the results of two experiments, one in which no microwaves were used and the delay was set to 10800 s, and a second in which the microwaves were turned on and the delay set to 5000 s.  The nutation pulse $\alpha$ was set to $\pi/2$ in both experiments.  The sign of the enhanced polarization is seen to be opposite to the thermal equilibrium polarization.  The observed enhancement of 20 corresponds to a silicon polarization of 0.68 \%.
        
Figure \ref{fig:power} shows the build up of the silicon  polarization under different conditions. Note that a 90 $\deg$ pulse is used to monitor the thermal silicon polarization in these experiments, while a 5 $\deg$ pulse is used to monitor the silicon polarization after DNP, which scales the signal by a factor of 11.5.  The original enhancement of 20 is increased by approximately a factor of 2 when we use a 1 W microwave source (Quinstar) instead of the 100 mW source \cite{Note1}.
%\footnote{The 1W source is obtained by injection-locking combining the outputs of two Impatt diode sources and injection locking them to a Gunn diode.}.
Reducing the temperature of the system from 1.4 K to 1.1 K increased the polarization by another factor of 1.3, proportional to the thermal electron polarization.  The $^{29}$Si signal is still increasing after 2 hours of microwave irradiation.  However the hold time of our cryostat reduced from 4 hours to about 2 hours with the increased microwave power and lower temperature.  Clearly an increased microwave irradiation time would allow us to achieve even higher nuclear spin polarizations.   The build up of the polarization could be fit quite well by a single exponential recovery curve.  We return to a discussion of the dynamics in section III.B.  

\begin{figure}
    \centering
    \includegraphics[width=5in]{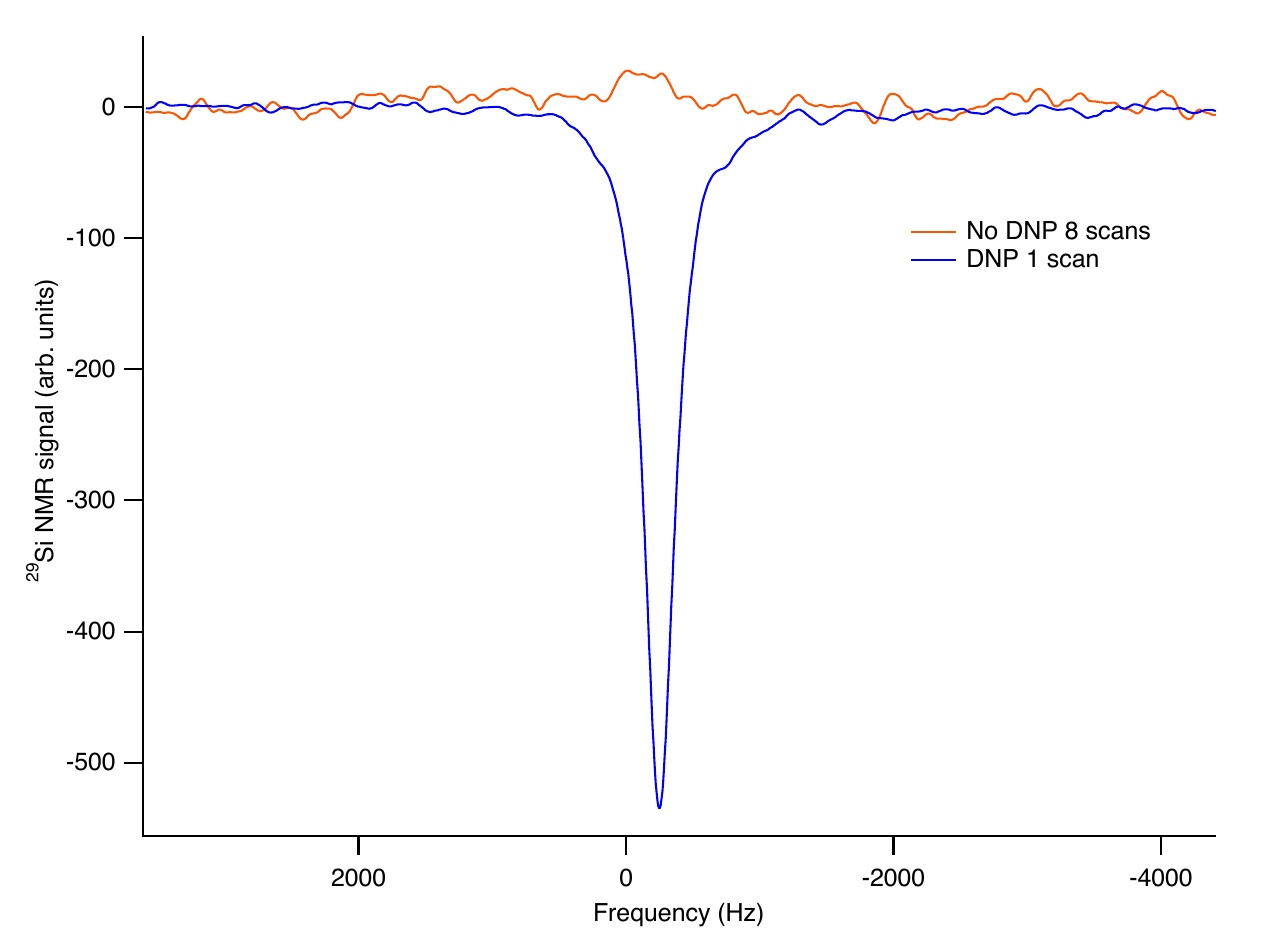}
    \caption{DNP of P-doped silicon (N$_D = 3 \times 10^{17}$ cm$^{-3}$) at 1.1 K.  The sample is placed in a tuned TE$_{011}$ cylindrical resonator during the microwave irradiation, and then physically moved approximately 1 inch to the center of  the RF coil where the polarization is measured.  The DNP signal (1 average) was recorded from a 2.8 mg piece of silicon wafer.  The thermal signal (8 averages) was recorded from a sample that also contained an additional 2.1 mg wafer fragment (total weight 4.9 mg).  The ratio of the measured signal intensities is $17.4\pm 5.8$, indicating a DNP enhancement of $244\pm81$ (color online). 
    }
    \label{fig:maxdnp}
\end{figure}

Since we were unable to further increase the power of our source, we replaced the microwave horn with a tuned cylindrical TE$_{011}$ cavity to increase the strength of the applied microwave magnetic field.  The RF coil was located just below the cavity, and the sample was physically moved from the cavity to the coil following microwave irradiation.  Since the T$_1$ of the silicon nuclei is on the order of hours, there is little loss of polarization during this process.  
At a temperature of 1.1 K we obtained a DNP enhancement of $244 \pm 81$ in phosphorus-doped silicon (N$_d\approx 2\times10^{17}$ cm$^{-3}$) after two hours of microwave irradiation,  which corresponds to a silicon nuclear spin polarization of $10.4 \pm 3.4$ \% (Figure~\ref{fig:maxdnp}), the highest value reported to date.  The thermal electron spin polarization is 90 \% under these conditions.  We were unable to measure a thermal silicon signal from the 2.8 mg piece of silicon wafer that was used in the DNP experiment.  A second 2.1 mg piece of wafer was added to the sample, and the thermal signal shown in Figure~\ref{fig:maxdnp} was recorded after 8 averages.  A $\pi/2$ nutation pulse was used in both experiments.  The ratio of the measured signal intensities is $17.4\pm 5.8$.  The low signal to noise ratio in this thermal signal is responsible for the relatively large uncertainty in the enhancement factor and the final silicon polarization.

\begin{figure}
    \centering
    \includegraphics[width=5in]{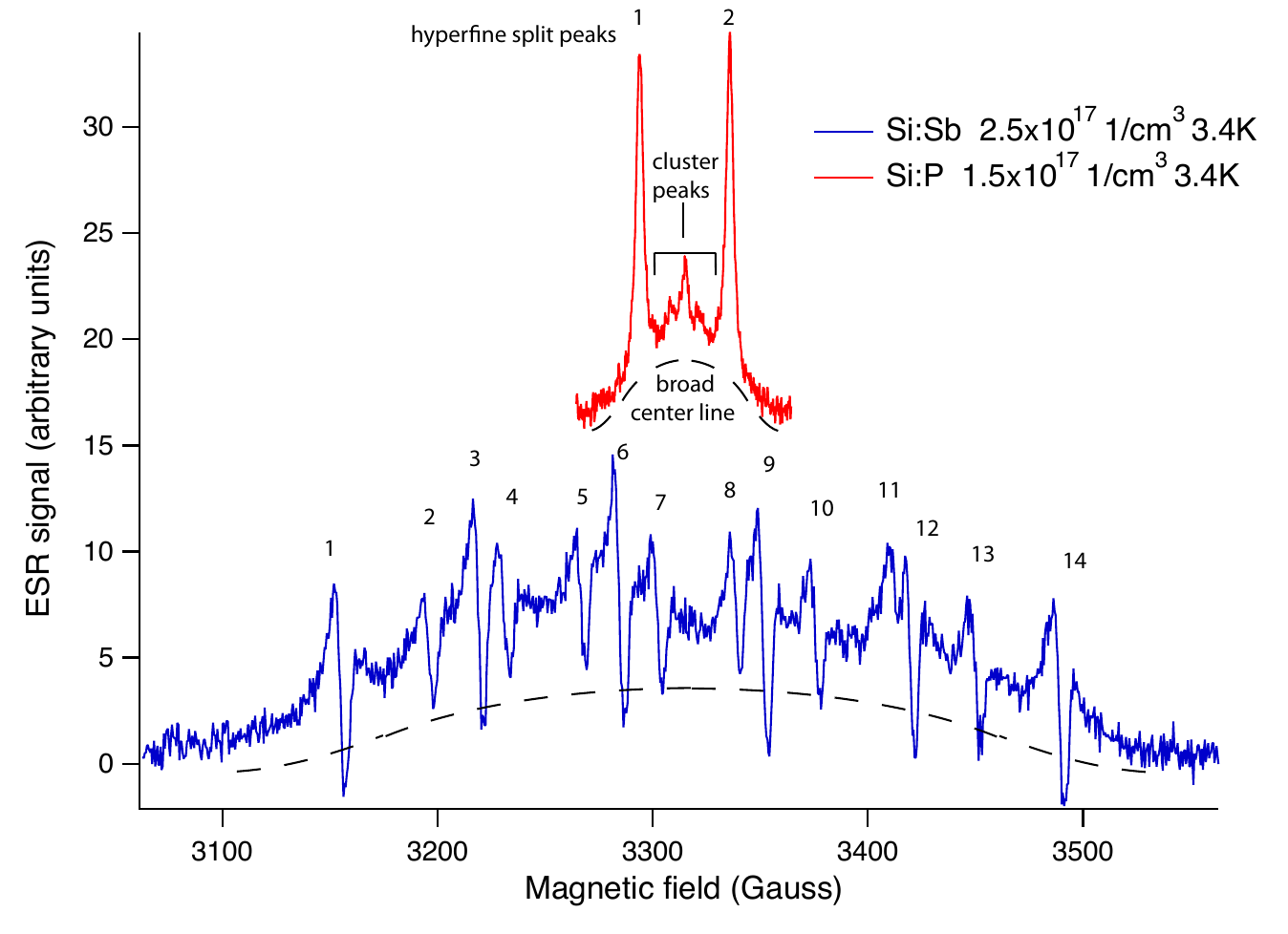}
    \caption{ESR spectra of phosphorus and antimony-doped silicon crystals obtained at X-band at a temperature of 3.4 K .  The signal from the phosphorus-doped sample was obtained under rapid-passage conditions.  The spectrum shows the two hyperfine resolved lines, the broad center line, and additional peaks that arise from donor clusters.   The signal from the antimony-doped sample did not quite satisfy the rapid passage conditions, resulting in some lineshape distortions.  The antimony doped sample shows the presence of 14 hyperfine resolved lines as well as the broad center line.  We did not resolve the cluster peaks in this sample. (color online)       }
    \label{fig:esr}
\end{figure}

\subsection{ESR Measurements}
\noindent In order to characterize our samples and further probe the DNP mechanism, we recorded ESR spectra from the samples and monitored the DNP signal as a function of the microwave irradiation frequency. 
Figure \ref{fig:esr} shows the results of low temperature (3.4 K) X-band ESR experiments on two of the samples studied. Given the random distribution of donors, the sample contains isolated donors as well as clusters of different sizes with an almost continuous distribution of exchange interaction strengths.  The measured ESR signal is an incoherent sum of these different contributions \cite{Cullis-1975}.  The spectra in Figure \ref{fig:esr} show several characteristic features of this sum.   While the signal from the phosphorus-doped sample was obtained under rapid-passage conditions, the signal from the antimony-doped sample did not satisfy the rapid passage conditions, resulting in some lineshape distortions.

\subsubsection*{ESR Spectrum}
\noindent Phosphorus-31 is 100 \% abundant and is a spin-1/2 nucleus with a donor hyperfine interaction of 117.5 MHz in silicon. The two isotopes of antimony --- $^{121}$Sb and $^{123}$Sb --- are 57.4 \% and 42.6 \% abundant, have nuclear spin 5/2 and 7/2 and hyperfine interaction strengths of 186.8 MHz and 101.5 MHz respectively \cite{Feher-1959}.  As a result the spectra from the isolated donors show 2 resolved hyperfine peaks for phosphorus and 14 for antimony.  In the phosphorus-doped sample, the small additional peaks in the spectra arise from exchange coupled clusters in which the exchange energy is much stronger than the donor hyperfine interaction \cite{Slichter-1955}.  If there are N strongly coupled donors the hyperfine interaction in the symmetric manifold of the donors is given by
\begin{equation}
\mathcal{H}_{hf}^S = \frac{A_D}{N}S_z \left(I_z^1 + I_z^2 + \ldots + I_z^N\right)
\end{equation}
where $A_D$ is the hyperfine interaction of a single donor.  
For a cluster containing N strongly coupled phosphorus donors (spin 1/2), there are N+1 resolved ESR transitions.  If the nuclear spins are unpolarized, the line intensities should follow a binomial distribution.  For strongly-coupled antimony clusters the spectra are much more complex given the presence of the two isotopes. These give rise to multinomial distributions.  For example, for a pair of spin-5/2 $^{121}$Sb, we expect 21 lines, which would follow a triangular distribution for unpolarized nuclei.  We did not resolve these peaks in the antimony-doped sample.  Finally both spectra show the presence of a broad background signal spanning the width of the resolved hyperfine interactions.  This background arises from exchange coupled clusters of two or more donors, where the exchange coupling is comparable to the donor hyperfine interaction \cite{Jerome-1964,Shimizu-1968,Cullis-1970}.  The width of the background is about 60 G for the phosphorus-doped sample and about 350 G for the antimony doped sample, corresponding to 170 MHz and 1 GHz respectively for $g \approx 2$.  The hyperfine coupling to the $^{29}$Si nuclei is not resolved and contributes an inhomogeneous linewidth of about 2.5 G to the spectra \cite{Feher-1959}.

The electron T$_1$ is concentration dependent in these samples.  
It is the onset of exchange-mediated effects that is responsible for the dramatic change in T$_1$ with doping concentration in the intermediate doping regime.  Isolated donors relax via spin diffusion to fast relaxing centers in the lattice, while exchange coupled pairs and clusters of spins relax to the lattice via the exchange reservoir \cite{Yang-1968}.

\begin{figure}
    \centering
    \includegraphics[width=5in]{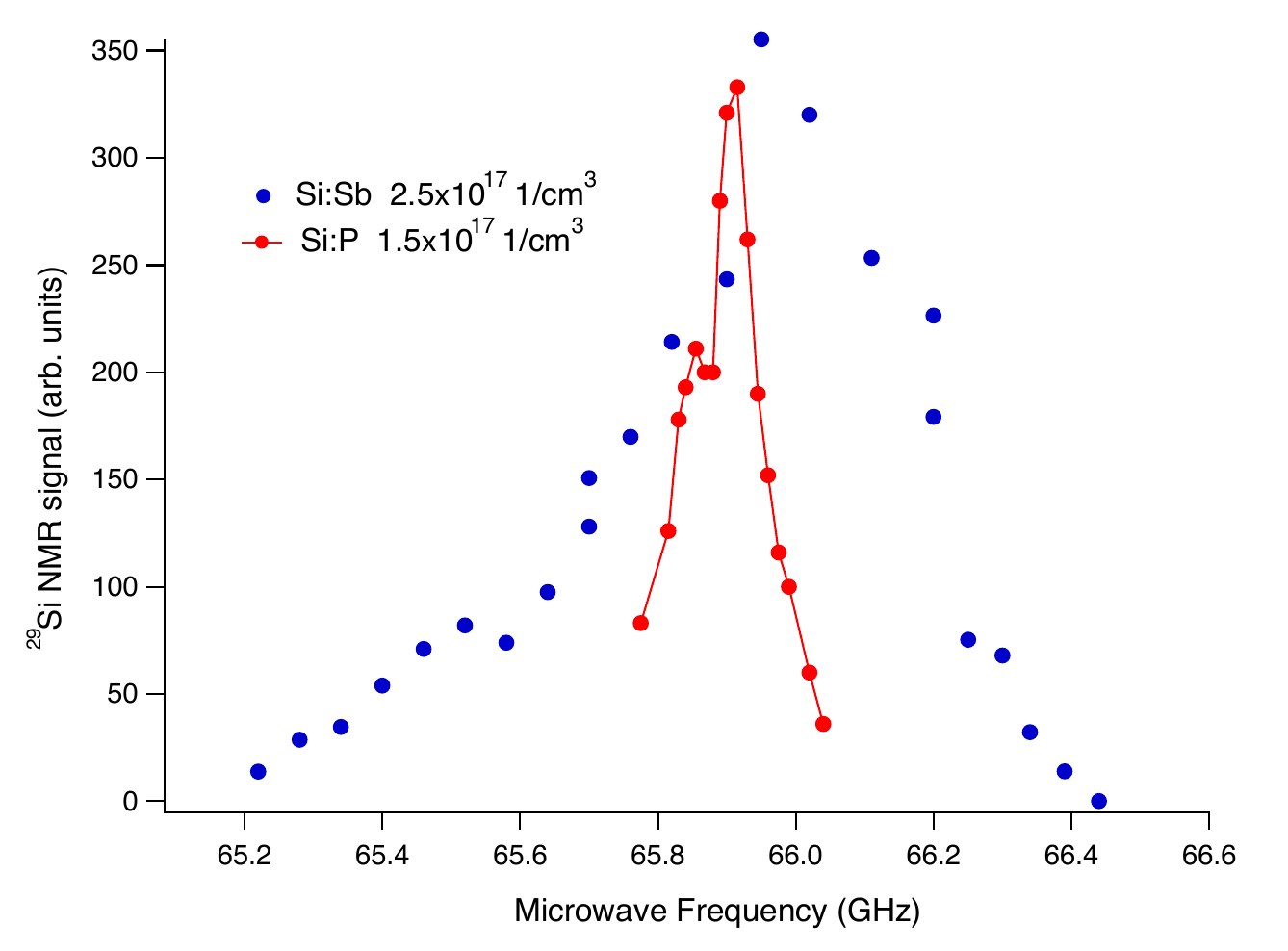}
    \caption{Frequency dependence of DNP enhancement for both phosphorus- and antimony-doped silicon, acquired with the 100 mW Gunn source (color online). }
    \label{fig:dnpfreq}
\end{figure}

\subsection{DNP Frequency Dependence}
\noindent Figure \ref{fig:dnpfreq} shows the DNP signal  (amplitude normalized) as a function of the microwave irradiation frequency.  The data were acquired with the 100 mW Gunn diode source.  DNP enhancement is observed over a frequency range of about 200 MHz for the phosphorus-doped silicon and about 1 GHz for the antimony-doped silicon, which is similar to the widths of the broad center lines in the ESR spectra in Figure \ref{fig:esr}.  The figures do not show any features at the frequencies of the donor-resolved hyperfine interactions for either sample.  
We do not see regions of positive and negative enhancement that are present in solid-effect DNP.
The sign of the enhancement does not change as we vary the microwave frequency, suggesting that an Overhauser mechanism is responsible for the enhancement.  The maximum Overhauser enhancement is given by $-\gamma_e / \gamma_n$.  Since both the electron and the silicon nucleus have negative gyromagnetic ratios, while the phosphorus and antimony nuclei have positive gyromagnetic ratios, the negative sign of the enhancement indicates a direct Overhauser enhancement of the silicon nuclei in both the phosporus- and antimony-doped samples.  

This suggests that the observed DNP is dominated by Overhauser enhancement of silicon nuclei within exchange-coupled clusters of donors, where the strength of the exchange coupling is comparable to the donor hyperfine coupling.  It is within these clusters that the electron spin T$_1$ become short enough to significantly modulate the contact hyperfine interaction and open up electron-nuclear cross-relaxation pathways.  The physics underlying the DNP process is the same for both phosphorus and antimony doped samples.  Both are shallow donors and the electron Bohr radius ($\sim 2$ nm) is expected to be very similar.  Thus, the strength of the exchange interaction depends on the spatial distribution of donors, which in turn depends on the donor concentration.  The donor hyperfine interaction and the exchange energy of the cluster are important in determining the frequencies of the allowed ESR transitions which is discussed in the following section.
Overhauser DNP has previously been observed in some charcoals in the presence of strong electron exchange interactions, where the ESR spectrum collapses to a single exchange-narrowed line \cite{Abragam-1958b}, but we believe this is the first time such an Overhauser effect has been observed in the presence of weaker exchange coupling strengths {\em at high field.}

\begin{figure}
    \centering
    \includegraphics[width=4in]{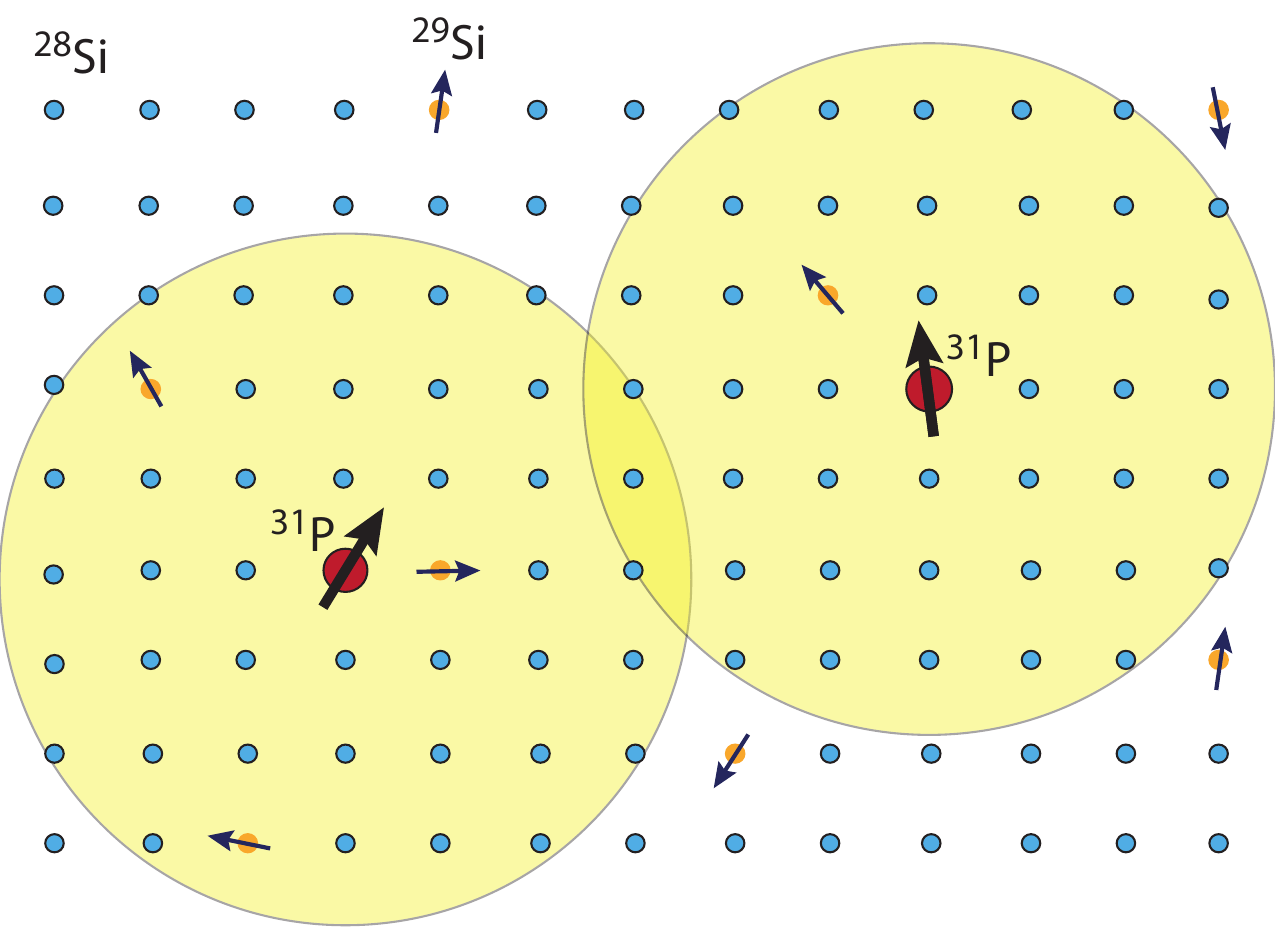}
    \caption{Schematic illustration of a two-donor cluster, showing the overlap of the electron orbitals with a number of $^{29}$Si nuclei (color online). }
    \label{fig:donors}
\end{figure}

\begin{figure}
    \includegraphics[width=5.6in]{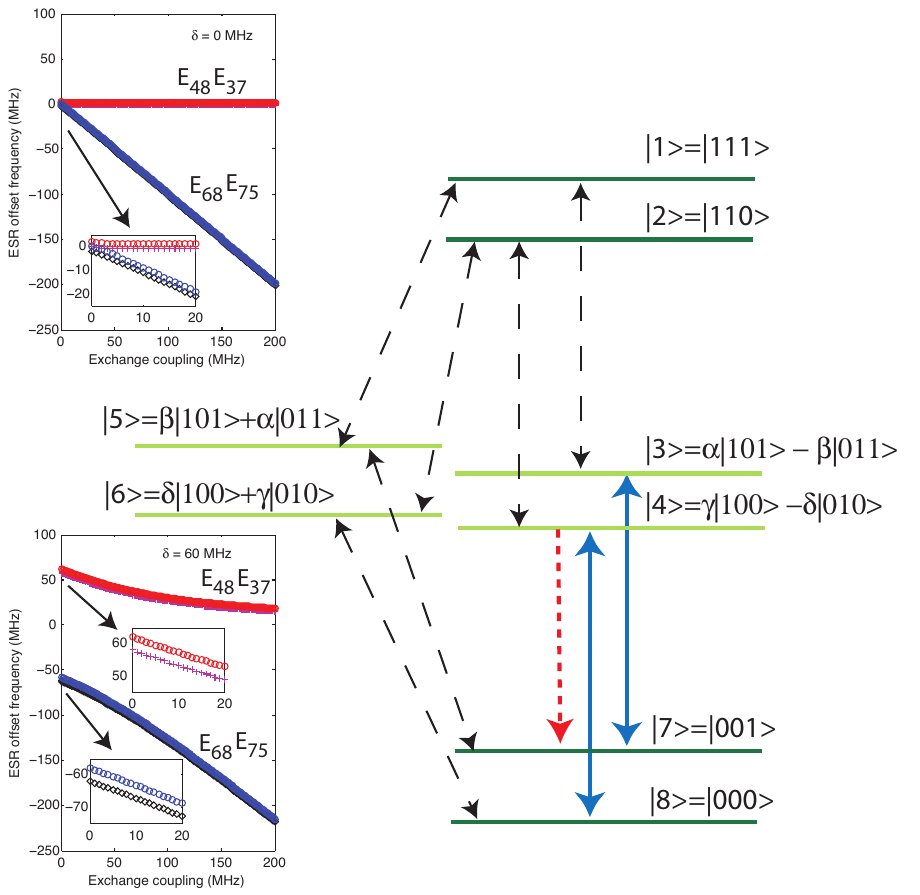}
    \caption{Eigenstates of the 3-spin Hamiltonian shown in Equation \ref{eq:3spin}, where $\alpha = \cos\frac{\theta}{2}$, $\beta = \sin\frac{\theta}{2}$,$\gamma = \cos\frac{\phi}{2}$, $\delta = \sin\frac{\phi}{2}$, $\tan\phi = \frac{-J_{\perp}}{2\delta + A/2}$, and $\tan\theta = \frac{-J_{\perp}}{2\delta - A/2}$.
The dashed black lines show the allowed ESR transitions that are not excited by the applied microwaves.  The solid blue lines show the allowed ESR transitions that are resonant with the applied microwave irradiation.  The dotted red line indicates the hyperfine mediated cross-relaxation path leading to DNP.  The energy levels shown in dark green correspond to the symmetric manifold while the levels shown in light green correspond to the asymmetric manifold.  We have labeled the states such that S$_z |0\rangle = -\frac{1}{2}|0\rangle$, and S$_z |1\rangle = \frac{1}{2}|0\rangle$.  The figures also shows the offset ESR frequencies $\omega-\omega_e$ as a function of the 
exchange coupling for the lower electron spin manifolds for $\delta = 0$ and $60$ MHz.  For $\delta = -60$ MHz the curves are almost identical to those for $\delta = +60$ MHz.  We used $A = 4$ MHz, $D = 0$ MHz, and $\omega_n = 20$ MHz.
(color online).}
    \label{fig:levels}
\end{figure} 

\subsection{Model}
\noindent Here we describe a simple  model to explain the frequency dependence of the Overhauser enhancement of the silicon nuclei in an exchange-coupled donor cluster (Figure~\ref{fig:donors}).  This also describes the frequencies of the ESR line. For simplicity we consider a cluster of phosphorus donors, since the nuclear spins have $I=1/2$.   The simplest model required would seem to be a 5 spin model --- containing two $^{31}$P donor nuclei, two electrons and a single $^{29}$Si nucleus that is located within the Bohr radius of one of the electrons.  We additionally assume that both the donor and the $^{29}$Si hyperfine interaction can be approximated by an isotropic Fermi-Segre contact interaction.  Since we only observe an Overhauser enhancement, we have also assumed that anisotropic hypefine interactions are negligible. The Hamiltonian of the 5 spin system is
\begin{eqnarray}
\mathcal{H} & = & -\omega_e\left(S_z^1 + S_z^2\right) + J\hat{S}^1\cdot\hat{S}^2 + D\left(2S_z^1S_z^2 - S_x^1S_x^2 - S_y^1S_y^2 \right)+  \nonumber \\
& & \hspace*{0.1in} A_D\left(S_z^1I_z^{D1} + S_z^2I_z^{D2}\right) + \omega_D\left(I_z^{D1} + I_z^{D2}\right) + AS_z^1I_z - \omega_nI_z
\end{eqnarray}
where $\omega_e$ is the electron Larmor frequency, $S^1$ and $S^2$ are the two electron spins, $I^{D1}$ and $I^{D2}$ are the donor nuclear spins, $I$ is the $^{29}$Si spins, $J$ is the exchange coupling strength, $D$ is the dipolar coupling between the electron spins, $A_D$ is the donor hyperfine coupling, $\omega_D$ is the Larmor frequency of the donor nuclei, $A$ is the hyperfine interaction with the $^{29}$Si nucleus and $\omega_n$ is the Larmor frequency of the silicon nucleus.  Note that $\omega_e$, $\omega_D$ and $\omega_n$ are all positive.  The exchange coupling is known to be antiferromagentic in silicon ($J > 0$) \cite{Jerome-1964}.  Here we have neglected the nuclear dipolar coupling as it is much weaker than any of the other interactions in the system.  It can be seen that the donor nuclei shift the energy levels, but always remain separable from the electrons.  If the donor nuclei are unpolarized, we have a distribution over the the corresponding electron spin energies.  This distribution narrows as the donor nuclei become polarized.
The $^{29}$Si spin also remains separable from the electronic spin states under this Hamiltonian.  We can reduce the dimensionality of the problem while capturing the essential physics if we replace the above Hamiltonian by a 3 spin Hamiltonian where the two electrons spins may be inequivalent (depending on the state of the associated $^{31}$P nuclear spins).
\begin{equation}
\mathcal{H} = \left(-\omega_e - \delta\right) S_z^1 + \left(-\omega_e + \delta\right) S_z^2 + J\hat{S^1}\cdot\hat{S^2} + D\left(2S_z^1S_z^2 - S_x^1S_x^2 - S_y^1S_y^2 \right) + AS_z^1I_z - \omega_nI_z
\end{equation}
which can in turn be re-written as
\begin{eqnarray}
\mathcal{H} & = & -\omega_e\left(S_z^1 + S_z^2\right) - \delta\left(S_z^1-S_z^2\right) + J_{||}S_z^1S_z^2 + \nonumber \\
& & \hspace*{0.1in} J_{\perp} \left(S_x^1S_x^2 + S_y^1S_y^2\right) + \frac{A}{2}\left(S_z^1+S_z^2\right)I_z + \nonumber \\
& & \hspace*{0.1in}  \frac{A}{2}\left(S_z^1-S_z^2\right)I_z - \omega_nI_z \label{eq:3spin}
\end{eqnarray}
where $J_{||} = J + 2D$ and $J_{\perp} = J - D$.
Since the electron Zeeman quantum number is a good quantum number, either the first or the second term is always zero, depending on the symmetry of the electron spin states.  In the symmetric manifold $S_z^1 = S_z^2 = \pm 1/2$ and $S_z^1 - S_z^2 = 0$, while in the antisymmetric manifold $S_z^1 = -S_z^2 = \pm 1/2$, and $S_z^1 + S_z^2 = 0$.  There are 4 levels in the symmetric manifold, two corresponding to $S_z^1+S_z^2 = 1$ and two corresponding to $S_z^1 + S_z^2 = -1$.    

The antisymmetric (or central) manifold has $S_z^1 + S_z^2 = 0$.  The structure of this manifold depends on the relative magnitudes of $A$,$\delta$ and $J_{\perp}$.  The eigenstates are shown in Figure \ref{fig:levels}.  The corresponding eigenenergies are  
\begin{equation}
\begin{array}{ll}
E_1 = \omega_e + \frac{\omega_n}{2} + \frac{A}{4} + \frac{J_{||}}{4}  &
E_2 = \omega_e - \frac{\omega_n}{2} - \frac{A}{4} + \frac{J_{||}}{4} \nonumber \\ 
E_3 = \frac{\omega_n}{2} + \frac{\omega_e^+ + \omega_e^-}{2} - \frac{J_{||}}{4} &
E_4 = -\frac{\omega_n}{2} + \frac{\omega_e^+ - \omega_e^-}{2} - \frac{J_{||}}{4} \nonumber \\ 
E_5 = \frac{\omega_n}{2} - \frac{\omega_e^+ + \omega_e^-}{2} - \frac{J_{||}}{4} &
E_6 = -\frac{\omega_n}{2} - \frac{\omega_e^+ - \omega_e^-}{2} - \frac{J_{||}}{4} \nonumber\\
E_7 = -\omega_e + \frac{\omega_n}{2} -\frac{A}{4} + \frac{J_{||}}{4} &
E_8 = -\omega_e - \frac{\omega_n}{2} + \frac{A}{4} + \frac{J_{||}}{4}
\end{array}
\end{equation}
where  $\omega_e^+ + \omega_e^- = \sqrt{\left(2\delta - \frac{A}{2}\right)^2 + J_{\perp}^2} $ and  $\omega_e^+ - \omega_e^- = \sqrt{\left(2\delta + \frac{A}{2}\right)^2 + J_{\perp}^2} $.

In our system $A \le 4$ MHz is the hyperfine coupling to the silicon nuclear spin.  The exchange coupling in different clusters range continuously from near zero to $J \approx 100$ GHz \cite{Cullis-1970}, and depending on the state of the phosphorus donor nuclei, $\delta$ takes on the value of  -60 MHz ($\downarrow\downarrow$), 0 MHz ($\downarrow\uparrow$ or $\uparrow\downarrow$) or 60 MHz ($\uparrow\uparrow$) which is half the hyperfine coupling to the phosphorus nuclei. The electron Zeeman frequency is 66 GHz.    In the limit that $J_{\perp} >> A, \delta$,  the two exchange coupled electrons form  singlet (levels 5 and 6) and  triplet (levels 1, 2, 3, 4, 7 and 8) manifolds as $\theta,\phi \rightarrow -\pi/2$, and transitions between these manifolds are forbidden.  This is why the width of the broad ESR line scales with the hyperfine coupling strength.  For intermediate values of the exchange coupling $J_{\perp} >> A$, $J_{\perp} \sim \delta$ microwave irradiation of the electron spins can induce the transitions $1\leftrightarrow3$, $1\leftrightarrow5$, $2\leftrightarrow4$, $2\leftrightarrow6$, $3\leftrightarrow7$, $5\leftrightarrow7$, $4\leftrightarrow8$ and $6\leftrightarrow8$ shown in Figure~\ref{fig:levels}.  The deviation of the ESR frequencies from the bare electron Larmor frequency $\omega - \omega_e$ are shown in the figure for the lower electron spin manifolds as a function of the strength of the exchange coupling. 

The transition efficiency depends both on the applied microwave power and how well the microwaves are tuned to the particular transition.  Since there is a random, quasi-continuous distribution of exchange coupling strengths in the sample, 
microwave irradiation at any frequency within the broad center line will be on-resonance for one of the transitions of an exchange-coupled pair.   For that pair, the microwaves will also be nearly on-resonance for a second transition as well (for example $4\leftrightarrow8$ and $3\leftrightarrow7$).  Since the strength of the applied B$_1$ field is on the order of a few MHz at most (1 W microwave power, low Q cavity), the other allowed transitions which are several tens of MHz off resonance are excited much less efficiently.  
In our experiments at 1.1 K and 2.35 T, only levels 7 and 8 are populated in thermal equilibrium, with almost equal populations, and the microwaves drives Rabi oscillations between the respective pairs of levels.  The exchange interaction between donor clusters also reduces the spin lattice relaxation times of the electron spins, and results in strong fluctuations of the local hyperfine interaction at the nuclear spin sites.  Essentially we are then dealing with just a simple 4 level system similar to that used in standard textbook descriptions of the Overhauser effect in liquids \cite{Abragam-1961}.  

We can consider these 4 levels (3,4,7 and 8 for example) represent a coupled spin-1/2 electron-nuclear system.  The Hamiltonian of the  system under microwave irradiation is given by 
\begin{equation}
\mathcal{H} = \tilde{\omega}_e S_z + \tilde{\omega}_n I_z + \tilde{A}I_zS_z + 2\omega_1 \cos \omega t S_x
\end{equation}
where
\begin{eqnarray}
\tilde{\omega}_e  & = & \omega_e + \frac{\omega_e^+}{2} -\frac{J_{||}}{2}  \\
\tilde{\omega}_n  & = & \omega_n + \frac{1}{2}\left(\omega_e^- - \frac{A}{2}\right)  \\
\tilde{A} & = & \omega_e^- + \frac{A}{2}  \: \: .
\end{eqnarray}

\subsection{DNP Dynamics}
\noindent If the microwaves are applied on-resonance $\omega = \tilde{\omega}_e$, the electron spins are driven into saturation as $\omega_1^2 T_1T_2 > 1$.  The degree of electron spin saturation is given by the saturation factor  
\begin{equation}
s = \frac{ S_0 - \langle S_z \rangle}{S_0} \: \: .
\end{equation}
The electron spin T$_1$'s of both the resolved hyperfine lines as well as the broad center line have been measured at X-band to be about 100 $\mu$s at 1.1 K  for P-doped Si with N$_D = 3 \times 10^{17}$ cm$^{-3}$, and is expected to be shorter in Sb-doped silicon.  
As the microwaves saturate the electron spin populations, the contact hyperfine interaction couples the relaxation dynamics of the two spin systems, and alters the populations of the nuclear spin.  The dynamics of the nuclear spin polarization are given by \cite{Abragam-1961}
\begin{equation}
\frac{d \langle I_z \rangle }{dt} =  -\frac{1}{T} \left\{ \langle I_z \rangle - I_0 \left(1-s\frac{\gamma_e}{\gamma_n}\right) \right \}
\label{eq:dnprate}
\end{equation}
where $I_0$ is the thermal equilibrium nuclear spin polarization, and
we have used the fact that $I=S=1/2$ and that the thermal electron spin polarization $S_0 = (\gamma_e/\gamma_n)I_0$.
The DNP time constant $T$ is given by \cite{Abragam-1961}
\begin{equation}
\frac{1}{T} = \frac{A^2}{2} \frac{\tau_2}{1 + (\omega_e-\omega_n)^2\tau_2^2} \approx \frac{A^2}{2\omega_e^2\tau_2} 
\label{eq:Tcp}
\end{equation}
where $A$ is the hyperfine coupling strength and $\tau_2$ is the correlation time of the transverse components of the electron spin.  
We have assumed here that there are no other sources of nuclear spin relaxation in these samples. 

There are two important time constants in the observed DNP of insulators.  The first is the time constant of the driven electron-nuclear interaction described above, and the second is due to nuclear spin diffusion \cite{Ramanathan-2008}.  As seen above $T$ depends on the strength of both the static external field and the applied microwave field, while the spin diffusion rate should be independent of external fields (as long as we remain in the high-field limit). In our experiments we observed only a single time constant in the growth of the DNP signal, which was observed to depend on the applied microwave power.  We therefore assume that the rate dynamics are limited by the Overhauser time constant and fit our data to the above model.  

Table~\ref{table1} shows the parameters obtained from fitting the data for antimony-doped silicon in Figure~\ref{fig:power} to Equation~\ref{eq:dnprate}, as well as the calculated saturation factor $s$ and the correlation time $\tau_2$ assuming $A \approx 1$~MHz.  Feher observed that the four most strongly-coupled silicon sites have hyperfine interactions ranging from 1--3 MHz for antimony-doped silicon, though the variation is not monotonic with distance in the immediate vicinity of the donor \cite{Feher-1959}.  At greater distances the interaction approximately falls off as $\exp(-r/r_0)$ where $r_0 \approx 2$ nm is the Bohr radius of the donor electron.  Thus the hyperfine coupling strength remains on the order of 1 MHz within one Bohr radius of the donor.  The saturation factors are quite low in the experiments shown in Figure 2.

In the absence of microwave irradiation $\tau_2^0 \approx 1.24$ $\mu$s.  This corresponds to just under a 1 MHz exchange interaction between clusters, which is the dominant source of these fluctuations, in agreement with the ESR and DNP experiments.  The correlation time is seen to get shorter as the microwaves power is increased.   Assuming that the microwave modulation of the spins and the exchange interaction are independent processes, we can estimate the strength of the applied microwave field using (see Table~\ref{table1})
\begin{equation}
\frac{1}{\tau_2} = \frac{1}{\tau_2^0} + f_1 \: \: \: .
\end{equation}

The strengths of the silicon hyperfine coupling and the donor exchange coupling do not change significantly in a similarly doped silicon sample.  We believe that the use of the cavity in the experiments shown in Figure~\ref{fig:maxdnp}, combined with the longer electron spin T$_1$ for P-doped silicon is responsible for the significant increase in the saturation factor and the observed DNP enhancement.  { In order to facilitate moving the sample from the cavity to the RF coil at low temperature, a hole was drilled in the side wall of the cylindrical resonator, significantly compromising its Q.  It should be possible to further increase the strength of the applied field with a higher Q cavity, and consequently improve the electron spin saturation and nuclear DNP enhancement.}

\begin{table}
\caption{Parameters obtained by fitting the data in Figure~\ref{fig:power} to Equation~\ref{eq:dnprate}, using $A = 2\pi \times 1$ MHz.  The amplitude of the DNP data was multiplied by 11.5 to account for the smaller flip angle pulsed used.}
\begin{tabular}{|c|r|c|c|c|c|c|} \hline
temp & \multicolumn{1}{c|}{T} & $I_0(1-s\gamma_e/\gamma_n)$ & $s$ & $\tau_2= A^2 T/ 2\omega_e^2$ & $f_1 $ & microwave source \\ \hline
1.4 K & 10839 s & $1.08 \times 10^5$ & 0 & 1.24 $\mu$s & 0 & none \\ 
1.4 K & 7593 s & -$1.78 \times 10^6$ & 0.005 & 0.87 $\mu$s & 0.3 MHz & 100 mW Gunn \\
1.4 K & 3400 s & -$2.79 \times 10^6$ & 0.008 & 0.39 $\mu$s & 1.8 MHz & 1 W Impatt \\
1.1 K & 3807 s & -$3.53 \times 10^6$ & 0.008 & 0.44 $\mu$s & 1.5 MHz & 1 W Impatt \\ \hline
\end{tabular}
\label{table1}
\end{table}

{ It is useful to re-examine our neglect of spin diffusion in the above calculations.
Assuming a uniform distribution of donors, the average distance between donors is on the order of 15 nm for a donor concentration of $3 \times 10^{17}$ cm$^{-3}$.  The hyperfine interaction of the donor electron with the silicon nuclei is on the order of 1--3 MHz for the first two shells of silicon nuclei where it is strongly anisotropic, and then decays exponentially on a characteristic length scale given by the Bohr radius.  In this case a 1 MHz hyperfine interaction would be reduced to 1 kHz at a distance of about 7 Bohr radii, suggesting that this represents the spin diffusion barrier around each donor.  However, the delocalized nature of the electronic wavefunction in these systems creates a hyperfine-mediated internuclear coupling that is typically stronger than the dipolar interaction between the spins \cite{Liu-2007}.  This coupling should lift the diffusion barrier, and provide an effective spin diffusion rate close to the dot that is faster than that due to dipolar couplings alone. 

The spin diffusion coefficient of natural abundance silicon is about $1 \times 10^{-14}$ cm$^2$/s or 1 nm$^{2}$/s ~\cite{Dementyev-2008,Hayashi-2009}.  Even for this diffusion rate, the time taken to transport the polarization a distance of a few nm --- required to polarize all the nuclear spins --- is on the order of a few tens of seconds which is much shorter than the timescales observed in the experiment.  This is in agreement with the results of Hayashi {et al}.\ who found that the spin-diffusion process was not a rate-limiting step in the low-field DNP of natural abundance silicon with N$_D$ in the range $10^{15}-10^{17}$ cm$^{-3}$ \cite{Hayashi-2009}.   At the lower doping concentrations the distance between donors becomes larger, and dipolar-mediated silicon nuclear spin diffusion will become important.
}

In summary, we have achieved high $^{29}$Si polarization in both phosphorus- and antimony-doped single crystal silicon. The  $^{29}$Si spins are directly polarized by donor electrons via an Overhauser mechanism within exchange-coupled donor clusters.  The Overhauser mechanism is observed even though the sample remains insulating at the low temperatures used in the experiment.  The physics underlying the DNP process is the same for both types of donors. Our results indicate that the key to achieving higher polarization in these samples is to improve the efficiency with which we saturate the electron spin polarization, as we did when using a resonant cavity for the microwaves.

\subsection*{Acknowledgments}
\noindent This work was funded in part by the National Security Agency under Army Research Office contract number W911NF0510469, the National Science Foundation under Award 0702295 and the Canada Excellence Research Chairs program. We thank Dr.\ Jonathan Hodges for several interesting discussions.

%\bibliography{Bibliography}


\end{document}